%% file: main.tex
\documentclass[12pt,3p]{elsarticle}

\usepackage{url}
\usepackage[T1]{fontenc}
\usepackage[utf8]{inputenc}
\usepackage{listings}
\usepackage{indentfirst}
\usepackage{subcaption}
\usepackage{lipsum}
\usepackage{multicol}
\usepackage{hyperref}
\usepackage{color, colortbl, soul}
\soulregister\cite7
\soulregister\ref7
\soulregister\pageref7
\soulregister\textit7
\definecolor{lightyellow}{rgb}{1,1,1}
\definecolor{lightGray}{gray}{0.9}
\sethlcolor{lightyellow}
\definecolor{yellow}{rgb}{1,1,1}
\usepackage{balance}
\usepackage{amssymb}
\usepackage{latexsym}
\usepackage{graphicx}
\usepackage{comment}
\usepackage{booktabs}
\usepackage{multirow}
\usepackage{framed}
\usepackage{comment}
\usepackage{longtable}
 \usepackage{float}
\usepackage{makecell}
\usepackage{soul}
\definecolor{Gray}{gray}{0.75}
\definecolor{LGray}{gray}{0.95}

\usepackage{comment}
\usepackage{mathtools}
\journal{Online Social Networks and Media}

\makeatletter
\def\ps@pprintTitle{%
   \let\@oddhead\@empty
   \let\@evenhead\@empty
   \let\@oddfoot\@empty
   \let\@evenfoot\@empty
}
\makeatother

\begin{document}

\begin{frontmatter}

 \title{A Behavioural Analysis of Credulous Twitter Users}

\author[label1,label2]{Alessandro Balestrucci}
\address[label1]{Gran Sasso Science Institute, via M. Iacobucci 2, 67100 L’Aquila, Italy}
\cortext[cor1]{Corresponding author}

\ead{a.balestrucci@gssi.it}

\author[label2,label4]{Rocco De Nicola}
\address[label2]{IMT School for Advanced Studies Lucca, Piazza San Francesco 19, 55100 Lucca, Italy}
\address[label4]{CINI Cybersecurity Lab, Via Ariosto, 25, 00185 Roma, Italy}
\ead{rocco.denicola@imtlucca.it}

\author[label3,label2]{Marinella Petrocchi\corref{cor1}}
\address[label3]{Istituto di Informatica e Telematica - CNR, Via G. Moruzzi 1, 56124 Pisa, Italy}
\ead{m.petrocchi@iit.cnr.it}

\author[label1]{Catia Trubiani}
\ead{catia.trubiani@gssi.it}

\begin{abstract}
%
Thanks to platforms such as Twitter and Facebook, people can know facts and events that otherwise would have been silenced.
However, social media significantly contribute also to fast spreading biased and false news while targeting specific segments of the population.
We have seen how false information can be spread using automated accounts, known as bots.
%
%
Using Twitter as a benchmark, we investigate behavioural attitudes of so called `credulous' users, i.e., genuine accounts following many bots. %
Leveraging our previous work, where supervised learning is successfully applied to single out credulous users, we improve the classification task with a detailed features' analysis and provide evidence that  simple and lightweight features are crucial to detect such users.
Furthermore, we study the differences in the way credulous and not credulous users interact with bots and discover that credulous users tend to amplify more the content posted by bots and argue that their detection can be instrumental to get useful information on possible dissemination of spam content, propaganda, and, in general, little or no reliable information.
\end{abstract}


\end{frontmatter}

\input{chapters/1-intro}
\input{chapters/2-backG}

\input{chapters/3-FeaturesEval}
\input{chapters/4-BehavA_New}

\input{chapters/5bis-Discussion}
\input{chapters/5-RW-shortened}

\input{chapters/6-Concl}

\bibliographystyle{elsarticle-num}
\bibliography{biblio}
\end{document}

%% file: chapters/1-intro.tex
\section{Introduction}
\label{sec:intro}

Disruptive Innovation: two words that sum up the impact of social networks and media on people's everyday life. Crucial information can be disseminated to millions of people in a flash: critical data, such as real-time updates on basic events. Unfortunately, new technologies have not only revolutionized traditional sectors such as retail and advertising. As noticed by the nonprofit organization National Endowment for Democracy, they have been fertile and groundbreaking even on a much more slippery ground: that of misinformation, hoaxes and propaganda~\cite{NationalEndowment2019}.
According to the 2019 report `Weapons of mass distractions'~\cite{gangware2019weapons}, strategists of false news can exploit - at least - three significant vulnerabilities of the online information ecosystem:
i) the medium: the platforms on which fake news creep in and expand;
ii) the message: the information one wants to convey; 
iii) the audience: the readers who consume (and contributes to diffuse) the information. 

This work focuses on the last aspect, i.e., the audience. 
Online Social Media convey the information quickly and diffusely. They are `optimised' for posting and sharing catchy and sensationalist news. False messages go from deliberate lies to mislead users to 
biased
information, aiming at influencing communities and agendas. Whatever the strategy adopted for spreading false news (like supporting automatic accounts or using trolls to inflame crowds~\cite{ferrara2016rise,Cresci17paradigm,shao2018spread,luceri2020trolls}), this would not be effective if there was no audience willing to believe them\footnote{Online: `Americans may appreciate knowing when a news story is suspect, but more than a third will share that story anyway'. Source \url{https://www.stopfake.org/}. All URLs in this manuscript were last accessed on January 3, 2021.}. The quest for belonging to a community and getting reassuring answers, the adherence to one's viewpoint, native reluctance to change opinion~\cite{Walton07aquestion,Webster97} are key factors for people to contribute to the success of disinformation spreading~\cite{Waytz2017,keersmaecker2017}. 


Information spreading on Social Media is often corroborated by automated accounts, called bots, which are totally or partially controlled by computer algorithms. Designed to
mimic human behaviour online, a dominant and worrisome use of automated accounts is far from being benign: they have been often used to amplify
narratives or drown out political dissent, see Ferrara et al. in~\cite{Ferrara2019arming}. Recent studies, such as the one by Shao et al.~\cite{shao2018spread}, demonstrate that bots are particularly active in spreading low credibility content. 
Moreover, the Oxford Internet Institute has monitored the global organization of social media manipulation by governments and political parties and analysed the trends of computational propaganda in different countries. The report~\cite{GlobalDisinformation2019} provides evidence of organized social media manipulation campaigns which have taken place in 70 countries, up from 48 countries in 2018 and 28 countries in 2017. In each country, there is at least one political party or government agency using social media to shape public attitudes domestically.

In our previous work~\cite{Balestrucci19ideal}, starting from the consideration that
human-operated accounts are exposed to manipulation and contribute to misinformation spreading by, e.g., retweeting or liking low-reputable content~\cite{shao2018spread}, we concentrated on Twitter and 
developed a classification framework for automatically detecting genuine accounts with a significant number of bots as friends, without exploiting this last datum. Hereafter, we define those users following many bots as `credulous' users.
This manuscript extends our previous work~\cite{Balestrucci19ideal} by analyzing the behavioural aspects (i.e., ratio on pure tweets, retweets, and replies) of credulous users, with the goal of understanding their distinguishing features.





The main goals of our research are summarized in the following:
(i) automatically identify genuine online users who may be prey of disinformation;
(ii) reduce misleading activities (e.g., spreading of fake news) performed by malicious entities like social bots;
(iii) stimulate users to verify the source of an information and fact-check the information itself, to pave the way to awareness.
    
To achieve these goals, we apply automated techniques to discriminate potential susceptible audiences and the accounts they interact with. 
%
%
%
Thus, the following four research objectives have been defined.
%
%
\begin{enumerate}
\item Assessing users' gullibility level: We propose
a technique to automatically rank human-operated accounts to assess their gullibility, by relying on aspects exposed on their social profiles. For instance, the number of bot accounts among their friends, or the number of by-bot-posts liked by the genuine user. 
%
\item  Detecting credulous users: We design and develop 
a supervised learning based classification framework to recognize those human-operated accounts following a high amount of bots. 
%
\item Profiling credulous users: 
We study the behavioral characteristics typical of credulous users, by analysing the interactions with their social contacts, and assessment of behavioral differences between credulous and not credulous users.
%
\end{enumerate}

The novel contributions of this manuscript, with respect to our previous work~\cite{Balestrucci19ideal}, are:
\begin{enumerate}
\item A deeper study of the features of the credulous classifier, with a specific analysis assessing the relevance of each single feature. 
\item An investigation of online behavior of users, in terms of tweets, retweets and replies, to better discriminate between credulous and non-credulous ones. 
\item A study of the actual influence of bots on credulous users by considering which and how many of the activities of credulous users are linked to tweets produced by bots.
%

\end{enumerate}


We can safely state that there exists a clear connection between the fact that a user has many bots among her/his friends and her/his actual contribution to amplifying the bots' messages. 
In particular, in this study we show that: 
\begin{enumerate}
    \item Lightweight features, such as the number of followers, tweets, and friends, are statistically significant to single out users with a high number of bots among their  friends;
    \item The `social activity reactions' to content originating from bots of credulous Twitter users is higher than that of not credulous users.
    \end{enumerate}
The experimental results are supported by statistical tests. 
We think that our methodology, which classifies credulous users with easy-to-extract features, is a promising new tool for finding content originating by automated accounts and, thus, for detecting spam, misleading information, and propaganda news. 




The remainder of the paper is organized as follows. 
Section~\ref{sec:background} briefly sums up our previous work. 
Section~\ref{sec:features} provides a study of the relevance of the features  exploited for the credulous users' classifier. Section~\ref{sec:behavior} discusses the behavior of credulous {\it vs} non-credulous users in terms of retweets and replies and  provides a fine-grained analysis of the extent to which  retweets and replies refer to tweets originated by bots. In Section~\ref{sec:discussion}, we discuss the main findings and implications of this investigation. Section~\ref{sec:RW}
discusses recent related work, positioning our study among relevant state-of-the-art papers.
Finally, Section~\ref{sec:concl} draws conclusions and highlights promising directions for future research and experimentation. The data used in this study are publicly available for the sake of reproducibility\footnote{\url{https://tinyurl.com/y6p7n38x}}.

%% file: chapters/2-backG.tex
\section{Background}
\label{sec:background}


In the following, we introduce some background notions reported in our previous work~\cite{Balestrucci19ideal,balestrucci2019credulous}, and present some of the performed experiments and the main findings. 
The main aim of this section is to provide a connection between what we have previously achieved and the analyses/experiments described in the following sections. 
Specifically, Section~\ref{subsec:datasets} introduces our datasets.  Section~\ref{subsec:botdetection} shows an excerpt of the experimental results related to the training of some bot detectors that we use to obtain the data used for the subsequent analyses. 
Section~\ref{subsec:CredClass} briefly describes the methodology applied for the identification of the credulous users 
and an excerpt of the experimental results, related to the training of credulous users detectors. 

\subsection{Datasets}
\label{subsec:datasets}
We considered three publicly available datasets: \texttt{CR15}~\cite{Cresci15fame}, \texttt{CR17}~\cite{Cresci17paradigm} and \texttt{VR17}~\cite{varol2017online}, 
where Twitter accounts are labelled according to their nature (either bots or not)\footnote{Bot Repository Datasets: \url{https://goo.gl/87Kzcr}}. 
%
\begin{itemize}
    \item[\texttt{CR15}:]  introduced in~\cite{Cresci15fame} consists of three smaller datasets.
The first one has been collected over a period of twelve days in December 2012, and contains 
469 Twitter accounts certified of being human-operated. The second one was collected between 2013-2015 and contains 1,488 genuine (human) users. The third subset is composed of 833 fake accounts, bought from three different Twitter accounts online markets.
%
 \item[\texttt{CR17}:] first presented in~\cite{Cresci17paradigm},  
 was obtained by following a hybrid crowd-sensing approach~\cite{avvenuti2017}. The authors 
 randomly contacted Twitter users by asking simple questions. All the replies were manually verified and 3,474 Twitter accounts were certified as humans. The dataset contains also 6,609 social spambots (e.g., spammers of job offers, products on sale at Amazon). 
 \item[\texttt{VR17}:] introduced in~\cite{varol2017online}, contains 2,573 Twitter accounts. 
 A manual annotation was performed by inspecting the profile details and the produced content. Overall, 1,747 Twitter accounts were annotated as human-operated and 826 as bots.
 \end{itemize}
%
%
%
From the merging of these three datasets, we obtain a unique labelled dataset (human-operated accounts/bots) of 12,961 accounts - 7,165 bots and 5,796 human-operated ones.

\subsection{Bot detection}
\label{subsec:botdetection}

The merged dataset was used to train a bot detector. To this end, we used the Java Twitter API\footnote{Twitter API: \url{https://goo.gl/njcjr1}}, and, for each account in the dataset, we collected: tweets (up to 3,200), mentions (up to 100) and IDs of friends and followers (up to 5,000).

In~\cite{Balestrucci19ideal}, we considered two features' sets derived from~\cite{varol2017online}\footnote{\url{https://botometer.iuni.iu.edu/}} and 
~\cite{Cresci15fame}. 
In particular, we relied on what Cresci at al. in ~\cite{Cresci15fame} called  \textit{ClassA} features, which conveniently require   only  information available in the profile of the account. 
\input{tables/tabBotDetectionReduced}
In Table~\ref{table:botDet} we report the results of the 19 learning algorithms adopted in~\cite{Balestrucci19ideal} to train the bot detector (with a 10-fold cross validation).
There are three reasons behind the decision to consider the classifier trained with the \textit{ClassA} features: (i) the performance results were very similar to those achieved considering the Botometer features, (ii) the features engineering phase rely on users' profile data only\footnote{Data from a user profile: \url{https://tinyurl.com/y5s5kpuw}}, and (iii) with respect to \textit{Botometer} features, where their calculation requires a connection to a web service\footnote{Botometer web service: \url{https://tinyurl.com/yytf282s}}, \textit{ClassA}'s features can be computed in an autonomous fashion. The training was executed also by considering the \textit{Botometer}'s features and a union set of \textit{ClassA}'s and \textit{Botometer}'s features.
Experiments were performed with Weka~\cite{witten2016data}, 
and the complete experimentation results are publicly available: \url{https://tinyurl.com/y4l632g5}.

\subsection{Classification of Credulous Twitter Users}
\label{subsec:CredClass}

In~\cite{Balestrucci19ideal}, we built a decision model to automatically classify Twitter accounts as credulous or not. As ground-truth to train the learning model, we considered 316 accounts belonging to the initial set of 5,796 human-operated ones, as introduced above (Section~\ref{subsec:datasets}).  
Due to the rate limits of the Twitter APIs and to the huge amount of friends possibly belonging to the ~6,000 genuine accounts, we considered only those accounts with a list of friends lower than or equal to 400~\cite{balestrucci2019credulous}.  
This leads to a dataset of 2,838 human-operated accounts, 
and 316 users have been identified as credulous ones, according to the approach  in~\cite{balestrucci2019credulous}. 



\input{tables/tabCredClassReduced}

We experimented with the same learning algorithms and the same features' sets considered in Section~\ref{subsec:botdetection}, with  a 10 cross-fold validation. 
It is worth noting that, for credulous users classification, the learning algorithms took as input a very unbalanced dataset: we had 2,838 human-operated accounts 
and, among them, 316 have been identified as credulous users.
To avoid working with unbalanced datasets, we split the sets of not credulous users into smaller portions, equal to the number of credulous users.
We randomly selected a number of not credulous users equal to the number of credulous ones; then, we unified these instances in a new dataset (hereinafter referred to as \textit{fold}). Then, we repeated this process on previously un-selected sets, until there were no more not credulous instances. Such procedure has been inspired by the \textit{under-sampling iteration} methodology, for strongly unbalanced datasets~\cite{lee2015iterative}.
Each learning algorithm was trained on each fold. 
To evaluate the classification performances on the whole dataset, and not just on individual folds, we computed the average of the single performance values, for each evaluation metric. 
Table~\ref{table:CredClassifiers} reports 
the classification performances  for the credulous users classifiers, obtained by using 19 learning algorithms. 
Also in this case, we used Weka to perform the experiments; further details are available here: \url{https://tinyurl.com/y4l632g5}.

%% file: tables/tabBotDetectionReduced.tex
\begin{center}
\begin{longtable}[htbp]
{p{0.1cm} p{1.0cm} p{1.7cm} p{1.7cm} p{1.5cm} p{1.5cm} p{1.5cm}}
& \multicolumn{5}{c}{\textit{evaluation metrics}}\\
\cline{3-7}
& \textit{alg} & \multicolumn{1}{r}{\textit{accuracy}} & \multicolumn{1}{r}{\textit{precision}} & \multicolumn{1}{r}{\textit{recall}} & \multicolumn{1}{r}{\textit{F1}} & \multicolumn{1}{r}{\textit{AUC}} \\
\hline
& HMM & \multicolumn{1}{r}{55.28} & \multicolumn{1}{r}{0.55} & \multicolumn{1}{r}{1.00} & \multicolumn{1}{r}{0.71} & \multicolumn{1}{r}{0.50} \\
& IBk & \multicolumn{1}{r}{91.03} & \multicolumn{1}{r}{0.91} & \multicolumn{1}{r}{0.93} & \multicolumn{1}{r}{0.92} & \multicolumn{1}{r}{0.91} \\
& BN & \multicolumn{1}{r}{87.15} & \multicolumn{1}{r}{0.93} & \multicolumn{1}{r}{0.83} & \multicolumn{1}{r}{0.88} & \multicolumn{1}{r}{0.94} \\
& NB & \multicolumn{1}{r}{64.37} & \multicolumn{1}{r}{0.89} & \multicolumn{1}{r}{0.42} & \multicolumn{1}{r}{0.54} & \multicolumn{1}{r}{0.77} \\
& VP & \multicolumn{1}{r}{80.07} & \multicolumn{1}{r}{0.82} & \multicolumn{1}{r}{0.82} & \multicolumn{1}{r}{0.82} & \multicolumn{1}{r}{0.80} \\
& MLP & \multicolumn{1}{r}{85.01} & \multicolumn{1}{r}{0.89} & \multicolumn{1}{r}{0.84} & \multicolumn{1}{r}{0.86} & \multicolumn{1}{r}{0.91} \\
& SMO & \multicolumn{1}{r}{68.58} & \multicolumn{1}{r}{0.76} & \multicolumn{1}{r}{0.63} & \multicolumn{1}{r}{0.69} & \multicolumn{1}{r}{0.69} \\
& JRip & \multicolumn{1}{r}{94.38} & \multicolumn{1}{r}{0.96} & \multicolumn{1}{r}{0.94} & \multicolumn{1}{r}{0.95} & \multicolumn{1}{r}{0.96} \\
& 1R & \multicolumn{1}{r}{84.51} & \multicolumn{1}{r}{0.88} & \multicolumn{1}{r}{0.84} & \multicolumn{1}{r}{0.86} & \multicolumn{1}{r}{0.85} \\
& 0R & \multicolumn{1}{r}{55.28} & \multicolumn{1}{r}{0.55} & \multicolumn{1}{r}{1.00} & \multicolumn{1}{r}{0.71} & \multicolumn{1}{r}{0.50} \\
& J48 & \multicolumn{1}{r}{94.30} & \multicolumn{1}{r}{0.96} & \multicolumn{1}{r}{0.94} & \multicolumn{1}{r}{0.95} & \multicolumn{1}{r}{0.96} \\
& HT & \multicolumn{1}{r}{84.48} & \multicolumn{1}{r}{0.90} & \multicolumn{1}{r}{0.81} & \multicolumn{1}{r}{0.85} & \multicolumn{1}{r}{0.88} \\
& RT & \multicolumn{1}{r}{92.48} & \multicolumn{1}{r}{0.93} & \multicolumn{1}{r}{0.94} & \multicolumn{1}{r}{0.93} & \multicolumn{1}{r}{0.92} \\
& J48c & \multicolumn{1}{r}{94.36} & \multicolumn{1}{r}{0.96} & \multicolumn{1}{r}{0.93} & \multicolumn{1}{r}{0.95} & \multicolumn{1}{r}{0.96} \\
& J48g & \multicolumn{1}{r}{94.41} & \multicolumn{1}{r}{0.96} & \multicolumn{1}{r}{0.94} & \multicolumn{1}{r}{0.95} & \multicolumn{1}{r}{0.96} \\
& LAD & \multicolumn{1}{r}{89.19} & \multicolumn{1}{r}{0.93} & \multicolumn{1}{r}{0.87} & \multicolumn{1}{r}{0.90} & \multicolumn{1}{r}{0.94} \\
& REP & \multicolumn{1}{r}{93.96} & \multicolumn{1}{r}{0.96} & \multicolumn{1}{r}{0.93} & \multicolumn{1}{r}{0.94} & \multicolumn{1}{r}{0.97} \\
& LMT & \multicolumn{1}{r}{94.33} & \multicolumn{1}{r}{0.96} & \multicolumn{1}{r}{0.94} & \multicolumn{1}{r}{0.95} & \multicolumn{1}{r}{0.97} \\
& \textbf{RF} & \multicolumn{1}{r}{\textbf{95.84}} & \multicolumn{1}{r}{\textbf{0.98}} & \multicolumn{1}{r}{\textbf{0.95}} & \multicolumn{1}{r}{\textbf{0.96}} & \multicolumn{1}{r}{\textbf{0.99}} \\
\hline
& & & & & & \\
\caption{Classification results for bot detection task with \textit{ClassA}'s features}
\label{table:botDet}
\end{longtable}
\end{center}

%% file: tables/tabCredClassReduced.tex
\begin{center}
\begin{longtable}[!t]
{p{0.1cm} p{1.0cm} p{1.7cm} p{1.7cm} p{1.5cm} p{1.5cm} p{1.5cm}}
& & \multicolumn{5}{c}{\textit{evaluation metrics}}\\
\cline{3-7}
& \textit{alg} & \multicolumn{1}{r}{\textit{accuracy}} & \multicolumn{1}{r}{\textit{precision}} & \multicolumn{1}{r}{\textit{recall}} & \multicolumn{1}{r}{\textit{F1}} & \multicolumn{1}{r}{\textit{AUC}} \\
\hline
& HMM & \multicolumn{1}{r}{50.06} & \multicolumn{1}{r}{0.50} & \multicolumn{1}{r}{1.00} & \multicolumn{1}{r}{0.67} & \multicolumn{1}{r}{0.50} \\
& IBk & \multicolumn{1}{r}{92.59} & \multicolumn{1}{r}{0.74} & \multicolumn{1}{r}{0.73} & \multicolumn{1}{r}{0.92} & \multicolumn{1}{r}{0.97} \\
& BN & \multicolumn{1}{r}{82.77} & \multicolumn{1}{r}{0.98} & \multicolumn{1}{r}{0.88} & \multicolumn{1}{r}{0.79} & \multicolumn{1}{r}{0.93} \\
& NB & \multicolumn{1}{r}{73.00} & \multicolumn{1}{r}{0.97} & \multicolumn{1}{r}{0.69} & \multicolumn{1}{r}{0.73} & \multicolumn{1}{r}{0.73} \\
& VP & \multicolumn{1}{r}{68.68} & \multicolumn{1}{r}{0.72} & \multicolumn{1}{r}{0.63} & \multicolumn{1}{r}{0.67} & \multicolumn{1}{r}{0.70} \\
& SMO & \multicolumn{1}{r}{75.32} & \multicolumn{1}{r}{0.74} & \multicolumn{1}{r}{0.80} & \multicolumn{1}{r}{0.77} & \multicolumn{1}{r}{0.75} \\
& MLP & \multicolumn{1}{r}{80.08} & \multicolumn{1}{r}{0.81} & \multicolumn{1}{r}{0.81} & \multicolumn{1}{r}{0.80} & \multicolumn{1}{r}{0.87} \\
& JRip & \multicolumn{1}{r}{93.05} & \multicolumn{1}{r}{0.99} & \multicolumn{1}{r}{0.87} & \multicolumn{1}{r}{0.93} & \multicolumn{1}{r}{0.94} \\
& 1R & \multicolumn{1}{r}{\textbf{93.27}} & \multicolumn{1}{r}{\textbf{0.99}} & \multicolumn{1}{r}{\textbf{0.88}} & \multicolumn{1}{r}{\textbf{0.93}} & \multicolumn{1}{r}{\textbf{0.93}} \\
& 0R & \multicolumn{1}{r}{49.51} & \multicolumn{1}{r}{0.49} & \multicolumn{1}{r}{0.65} & \multicolumn{1}{r}{0.66} & \multicolumn{1}{r}{0.50} \\
& J48 & \multicolumn{1}{r}{92.58} & \multicolumn{1}{r}{0.97} & \multicolumn{1}{r}{0.88} & \multicolumn{1}{r}{0.92} & \multicolumn{1}{r}{0.94} \\
& HT & \multicolumn{1}{r}{83.28} & \multicolumn{1}{r}{0.96} & \multicolumn{1}{r}{0.71} & \multicolumn{1}{r}{0.80} & \multicolumn{1}{r}{0.93} \\
& RT & \multicolumn{1}{r}{88.88} & \multicolumn{1}{r}{0.89} & \multicolumn{1}{r}{0.89} & \multicolumn{1}{r}{0.89} & \multicolumn{1}{r}{0.89} \\
& J48C & \multicolumn{1}{r}{92.68} & \multicolumn{1}{r}{0.97} & \multicolumn{1}{r}{0.88} & \multicolumn{1}{r}{0.92} & \multicolumn{1}{r}{0.94} \\
& J48g & \multicolumn{1}{r}{92.64} & \multicolumn{1}{r}{0.97} & \multicolumn{1}{r}{0.88} & \multicolumn{1}{r}{0.92} & \multicolumn{1}{r}{0.94} \\
& LAD & \multicolumn{1}{r}{92.38} & \multicolumn{1}{r}{0.96} & \multicolumn{1}{r}{0.89} & \multicolumn{1}{r}{0.92} & \multicolumn{1}{r}{0.97} \\
& LMT & \multicolumn{1}{r}{92.66} & \multicolumn{1}{r}{0.98} & \multicolumn{1}{r}{0.88} & \multicolumn{1}{r}{0.92} & \multicolumn{1}{r}{0.96} \\
& REP & \multicolumn{1}{r}{93.09} & \multicolumn{1}{r}{0.98} & \multicolumn{1}{r}{0.88} & \multicolumn{1}{r}{0.93} & \multicolumn{1}{r}{0.95} \\
& RF & \multicolumn{1}{r}{92.71} & \multicolumn{1}{r}{0.97} & \multicolumn{1}{r}{0.89} & \multicolumn{1}{r}{0.92} & \multicolumn{1}{r}{0.97} \\
\hline
& & & & & & \\
\caption{Classification results for credulous detection with \textit{ClassA}'s features.}
\label{table:CredClassifiers}
\end{longtable}
\end{center}

%% file: chapters/3-FeaturesEval.tex
\section{Features' Evaluation}
\label{sec:features}

The original contribution of this manuscript starts here.
We extend the credulous classification analysis to assign each \textit{ClassA}'s feature an `index of ability' to distinguish C from NC instances\footnote{In the following, we will adopt notation C and NC users to indicate, resp. credulous and not credulous accounts.}. Table~\ref{tab:ClassANot} presents the \textit{ClassA} features with their type and description.

\input{tables/tabFeatNotationWithoutType}

\vspace{-2em}
\subsection{Ranking of \textit{ClassA} features}
Weka's tools allow to assess the discriminatory importance of a feature in a features' set through the so called \textit{attribute selection}. For the sake of reliability, we consider three attribute selector algorithms that evaluate the value (in terms of importance) of each attribute with different methodologies:  (i) \textit{OneRAttributeEval}\footnote{OneRAttributeEval: \url{https://tinyurl.com/qtl3nox}} 
uses the OneR classifier, 
(ii) \textit{SymmetricalUncertAttributeEval}\footnote{SymmetricalUncertAttributeEval: \url{https://tinyurl.com/wcgccoz}} 
measures the symmetric uncertainty with respect to the class and (iii) \textit{InfoGainAttributeEval}\footnote{InfoGainAttributeEval: \url{https://tinyurl.com/ve99qt8}} 
considers the information gain~\cite{kent1983information} against the class. 

\input{tables/tabAttributeRanks}

Table~\ref{tab:FeatRanks} shows the ranking of the first six most important features,  
according to the three evaluating algorithms. 
The remaining features have been estimated to impact with a lower relevance, in fact at least one of the evaluators estimated a value lower than 0.1, this happens for the seventh feature in the rank (i.e., $F9$) estimated as follows: 0.631 (\textit{OneRAttributeEval}), 0.101 (\textit{SymmetricalUncertAttributeEval}) and 0.085 (\textit{InfoGainAttributeEval}).
%
From Table~\ref{tab:FeatRanks}, we can see that all the attribute evaluators confirm the relevance of the same features in the first six positions.

\subsection{Analysis on three specific features}
\label{subsec:ttestFeat}
Here, we carry out a further analysis on three specific features, F3 (\#tweets), F5 (\#friends) and F19 (\#followers). The rationale behind this feature selection is due to the following reasons:
(i) these features are direct and simple indicators of the level of the account's activity (F3) and friendliness (F5 and F19), (ii), they are not related between each other (like, for example, F1 and F7), and (iii) we think they are more easily understandable rather than a combination of the same, see F1 and F14. Furthermore, given the specific nature of the statistical tests carried on in the following, we do not consider boolean features.

The statistical tests 
are carried on to determine if the values of the three features are statistically significant to discriminate between C users and NC users. Precisely, a \textit{paired t-test}~\cite{hsu2005paired} (with $\alpha=0.05$) is a well known parametric statistical test where the observations of some values of two populations are compared; the goal is to verify whether the average values of the two value distributions significantly deviate between each other.
 Furthermore, the Pearson Correlation Coefficient (PCC) has been calculated to single out any correlation between each feature data value. The PCC is an index expressing a possible relationship of linearity between two statistical variables. PCC values are between +1 and -1, where +1 corresponds to the perfect positive linear correlation, 0 corresponds to an absence of linear correlation and -1 corresponds to the perfect negative linear correlation. 

\input{tables/tabStats}

Tests are carried out over 316 C users and 316 NC users. The 316 NC users have been randomly selected  (without re-entry) from the full set of 2.522 NC users. Table~\ref{tab:Ttest} shows the p-values (derived from the t-test) and the PCCs. Results have been obtained with the use of the \textit{commonMaths Apache's library}\footnote{Commons-Math library (Apache): \url{https://tinyurl.com/lt7zeud}}.

Seeing at the values in the table, we argue that the two populations (C and NC users) feature a difference, w.r.t. the considered features,  and this difference is statistically significant, since the p-values are practically equal to 0.
Also, the fact that PCC is, for the three features, very close to 0 implies that there is no linear correlation between the values, per feature, of the two populations. 

%% file: tables/tabFeatNotationWithoutType.tex
\begin{center}
    \begin{longtable}[htbp]
    {p{1cm} p{5cm} p{9cm}}
    \hline
    \textbf{Label} & \textbf{Feature Name}  & \textbf{Description}\\
    \hline
    &&\\
        F1 & \#friends/\#followers$^2$  & The ratio between the number of friends and the squared number of followers \\[3mm]
        F2 & age (in months)  & The number of months since the creation of the account\\[3mm]
        F3 & \#tweets  & The number of tweets, retweets, replies and quotes of the account\\ [3mm]
        F4 & has a Name  & True if a name is specified in the account's profile\\[3mm]
        F5 & \#friends  & (Alias \#followees): The number of accounts a user is following\\[3mm]
        F6 & URL in profile  & True if a URL is specified in the account's profile \\[3mm]
        F7 & following rate  & The number of followees over the sum of followees and followers\\[3mm]
        F8 & default image after 2m  & True if the account did not change the default image provided by Twitter in the account's profile after 2 months of its creation\\[3mm]
        F9 & belong to a list  & True if the account is member of, at least, one list \\[3mm]
        F10 & profile has image  & True if the account has an image in its profile\\[3mm]
        F11 & \#friends/\#followers $\geq$ 50  & True if the ratio between the number of friends and followers is greater than or equal 50\\[3mm]
        F12 & `\textit{bot}' in bio  & True if  there is a clear declaration of being a bot in the account's profile\\[3mm]
        F13 & duplicate profile pictures  & True if the profile's image is the same of that of other accounts  (We do not consider this feature in the current work) \\[3mm]
        F14 & 2 x \#followers $\geq$ \#friends  & True if twice the followers is greater than or equal the number of followees\\[3mm]
        F15 & \#friends/\#followers $\simeq$ 100 & True if  an account is following a number of accounts that is about 100 order of magnitude the number of accounts that follows it \\[3mm]
        F16 & profile has address  & True if a location is specified in the  account's profile\\[3mm]
        F17 & \multicolumn{1}{l}{\makecell{no bio, no location, \\ \#friends $\geq$ 100}}  & True if: the account has no description in the bio and location fields of its profile and the number of friends is greater than or equal 100 \\[3mm]
        F18 & has biography  & True if the biography is specified in the account's profile \\[3mm]
        F19 & \#followers  & The number of the account's followers\\[3mm]
        \hline
        & & \\
        \caption{Type and description of \textit{ClassA}'s features}
    \label{tab:ClassANot}
    \end{longtable}
\end{center}

%% file: tables/tabAttributeRanks.tex
\begin{table}[htbp]
    \centering
    \begin{tabular}{lccc}
    \hline
\textbf{Rank} &
\textbf{OneR} &
\textbf{SymmetricalUncert} &
\textbf{InfoGain}\\
\hline
1 & F1 (1.000) &  F1 (1.000) &  F1 (1.000) \\
2 & F14 (0.977) &  F14 (0.896) & F14 (0.894) \\
3 & F19 (0.889) &  F19 (0.509) & F19 (0.620) \\
4 & F3 (0.768) &  F5 (0.299) & F3 (0.323) \\
5 & F5 (0.720) &  F7 (0.235) & F5 (0.273) \\
6 & F7 (0.712) &  F3 (0.218) & F7 (0.255) \\
\hline
    \end{tabular}
    \caption{Ranking of relevance of ClassA features, according to the three attribute evaluators}
    \label{tab:FeatRanks}
\end{table}

%% file: tables/tabStats.tex
\begin{table}[ht]
    \centering
    \begin{tabular}{lccc}
    \hline
         & \textbf{F3} & \textbf{F5} & \textbf{F19} \\
         \hline
         \textit{P-value} & 6.211$\times 10^{-24}$ & 1.166$\times 10^{-34}$ & 5.005$\times 10^{-12}$ \\ 
         \textit{PCC}  & -0.019 & 0.061 & 0.001 \\
         \hline
    \end{tabular}
    \caption{Statistical significance test (T-test with $\alpha=0.05$) on F3, F5 and F19. 
    }
    \label{tab:Ttest}
\end{table}

%% file: chapters/4-BehavA_New.tex
\section{Behavioral Analysis}
\label{sec:behavior}

\begin{figure}[ht!] 
   \begin{subfigure}[b]{0.5\linewidth}
     \centering
     \includegraphics[width=0.85\linewidth]{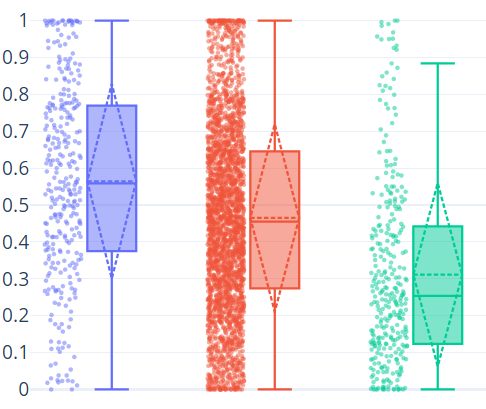} 
     \caption{Pure Tweets ratio} 
     \label{fig7:a} 
     \vspace{4ex}
   \end{subfigure}
   \begin{subfigure}[b]{0.5\linewidth}
     \centering
     \includegraphics[width=0.85\linewidth]{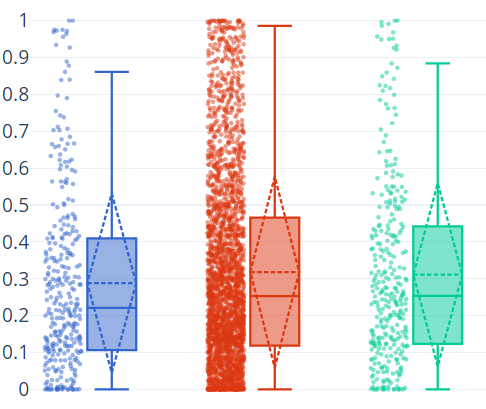} 
     \caption{Retweets ratio} 
     \label{fig7:b} 
     \vspace{4ex}
   \end{subfigure} 
\begin{center}
   \begin{subfigure}{0.5\linewidth}
     \includegraphics[width=0.85\linewidth]{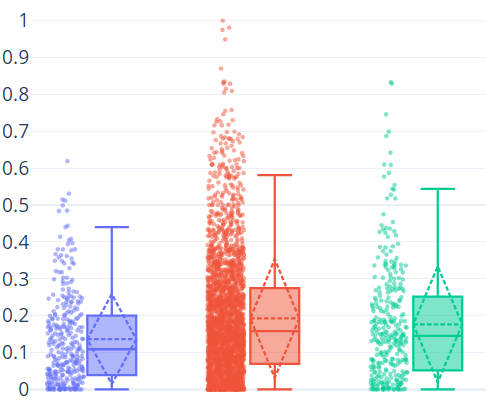} 
     \caption{Replies ratio} 
    \label{fig7:d} 
  \end{subfigure}
  \begin{subfigure}[r]{0.4\linewidth}
     \centering
     \includegraphics[width=0.8\linewidth]{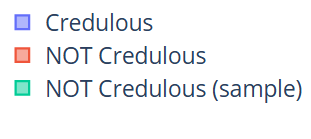} 

     \label{fig7:Leg} 
   \end{subfigure}\\
  \end{center}
  \caption{Activities of credulous users ({\it vs} not). Each plot expresses the ratio between the tweets in the user's timeline and (a) content produced by the user, (b) retweets, and (c) replies\label{fig:actions}}
 \end{figure}

 
In this section, we analyse the activities of credulous accounts, in terms of tweets (Figure~\ref{fig7:a}), \textit{retweets} (Figure~\ref{fig7:b}), and \textit{replies} (Figure~\ref{fig7:d}).  
Quoted tweets have been considered as retweets\footnote{On Twitter, a quoted tweet is a retweet with an extra text inserted by the retweeter.}.  Results are shown in  Figure~\ref{fig:actions}. For each type of content, each subfigure reports statistics about  users' activities for: 
%
the 316 C users (blue), the 2,522 NC users (red), and a random sample of NC users of the same number of C ones, 316 (green). 

Figure~\ref{fig7:a} reports  information related to pure tweets. When related to the overall amount of tweets, C users (blue points) produced, on average, 56.44\% of tweets (horizontal blue dashed line), with a standard deviation (dashed blue rhombus) of 26.4\%. The totality of NC users  (red points) feature an average tweets production that is lower than C users, precisely 46.49\% ($\sigma$=25.45\%). When considering the sample of NC users (green points), we notice an even lower average  (31.13\%, $\sigma$=24.85\%). The analysis of this first graph suggests that those accounts classified as credulous tweet 
original content more than the others.

Figure~\ref{fig7:b} reports the information related to retweets and quotes (w.r.t. the overall amount of tweets). In this case, the difference between C and NC users is less marked. C users (blue points) show a retweets-tweets ratio equal to 0.2882 ($\sigma$=0.2432), while NC users (red points) ratio is 0.3182 ($\sigma$=0.2591). Very similar scores are obtained if the NC users' sample (green points) is considered, with average ratio =0.311 ($\sigma$=0.2485). 

Similar findings have been obtained for the replies, see  Figure~\ref{fig7:d}. The replies-tweets ratio is equal to 0.14 ($\sigma$=0.124) for C users (blue points), on average. The same ratio for the NC population (red points) is higher, with a value equal to 0.19 ($\sigma$=0.164). For the NC users' sample, we obtain 0.18 ($\sigma$=0.16).

Although the last two cases (retweets and replies) show a common decreasing trend in the averages of the activities of C and NC users, we will further investigate the issue with a fine grained analysis, with the aim of finding more discriminant results in terms of C and NC users' behavior. 
%
Precisely, we will analyse the nature of the accounts that originated retweets and replies of C and NC users.
%
%
For each of the 2,838 human-operated accounts in our dataset, 
and for the two kind of actions type of action --retweeting and  replying --, we will calculate the percentage of content originated by bots. 
Considering, for example, the case of retweets, it is possible to retrieve the ID of the original tweet. Consequently, from the tweet ID, it is possible to retrieve the tweet author. We can then evaluate if that author is classified as bot or not. A similar reasoning holds for C users' replies (considering the nature of the author of the tweet which the user replies to) and quotes.

For the bot classification task, we adopt the classifier presented in Section~\ref{subsec:botdetection}. The authors of the original tweets retweeted and quoted by our human-operated accounts, or to which they responded, are 1.22 million users. Among them, 104k has been classified as bots (8.5\%).

\subsection{Retweets}
\label{subsec:fineRw}
\label{subsec:retweetsN}

\begin{figure}[ht!] 
\hspace{-4em}
   \begin{subfigure}[b]{0.6\linewidth}
     \centering
     \includegraphics[width=0.95\linewidth]{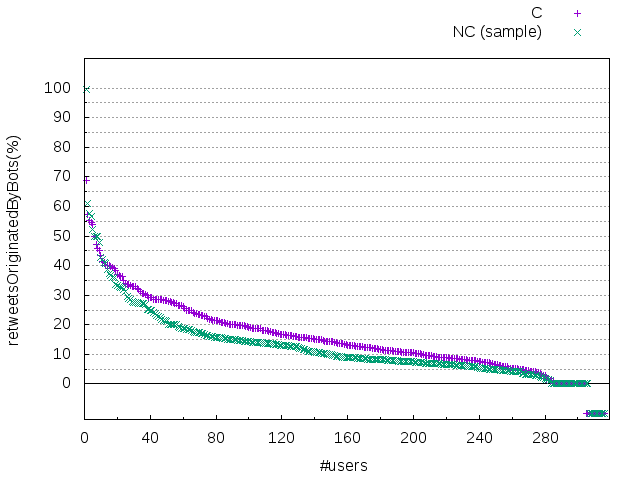} 
     \caption{Percentage of retweets originated by bots} 
     \label{fig:rwperc} 
   \end{subfigure}
   \hspace{-1em}
   \begin{subfigure}[b]{0.6\linewidth}
     \centering
     \includegraphics[width=0.95\linewidth]{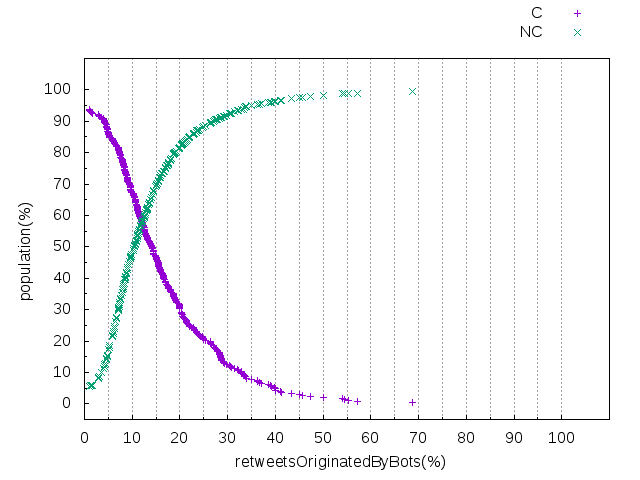} 
     \caption{\% of populations w.r.t. the \% of retweets originated by bots} 
     \label{fig:rwcov} 
   \end{subfigure}
\caption{Comparative analysis between \textit{credulous} and \textit{not credulous} users with respect to the retweets whose related tweets have been originated by bots.}
    \label{fig:rw}
\end{figure}

Figure~\ref{fig:rw} gives two different views of the same phenomenon.  In both subfigures, C users are represented in purple, while NC users in green.

Figure~\ref{fig:rwperc} gives, on the y-axis, the percentage of retweets whose original tweets have been originated by bots\footnote{For the sake of briefness, hereafter we will denote such retweets as `byBot-retweets'.}. Numbers on the x-axis are indexes for each user, instead of the Twitter ID; such values are useful to count the number of users with a percentage of byBot-retweets
greater than a certain threshold. 
%

It is worth reminding that the original NC set is composed of 2,522 users; hence, for sake of fair comparison, in the figure we consider a sample of 316 NC users.  To obtain a representative sample,  
we first built 20 samples of 316 NC users; each sample was obtained by randomly selecting instances from the original set, without re-injection. Then, for each sample, we computed the average and standard deviation on the percentage of byBot-retweets.  Finally, we computed the Euclidean distance between the averages and standard deviations of the samples and we compared them to those calculated over the entire NC population. We identified as more representative of the entire NC population the sample with the smallest distance. 


%
 Looking at Figure~\ref{fig:rwperc}, we can notice that the purple points (C users) are above the green ones (sample of NC users). 
 %
The average percentage of tweets originated by bots retweeted by C users is 16.45 ($\sigma=11.84$\%), while the average percentage for NC users is lower, 13.21 (with $\sigma=12.1$\%). 

The percentage of byBot-retweets  have been calculated over the total amount of retweets. Some of the human-operated accounts in our dataset do not retweet at all. We call such accounts outliers. In Figure~\ref{fig:rwperc}, the outliers are shown under the zero on the y-axis: 10 C users and 12 NC users are outliers. 
Moreover, the users lying exactly on the y-axis are those users who retweet only tweets  originated by human-operated accounts.



 
 Figure~\ref{fig:rwcov} compares the whole C and NC populations. The values on the x-axis are the same of those on y-axis in Figure~\ref{fig:rwperc}. Instead, on the y-axis, we report the percentage of the population having byBot-retweets  (in percentage) {\it greater or equal to} (for C users -- purple dots) or {\it lower} (for NC users -- green dots) to the values on the x-axis. 

The aim of the graphs in Figure~\ref{fig:rwcov} is conveying a measure of \textit{population coverage}, i.e., fixing the number of byBot-retweets, we know the percentage of C users whose byBot-retweets  is $\ge$ that number and the percentage of NC users which quotes is less than  that number. In Figure~\ref{fig:rwcov}, the data related to NC users refer to all of them (2,522).


%
The green and purple curves  reach the maximum population coverage (of users) at the abscissa point of 15.59 (\%). Specifically, the 43.75\% of C users has a percentage of byBot-retweets $\ge$ 15.59 (coordinates 15.59, 43.75 -- purple dots). The 70.04\% of NC users has a percentage of byBot-retweets  < 15.59 (coordinates 15.59, 70.04 -- green dots).

\begin{figure}[ht!] 
\begin{subfigure}[b]{1\linewidth}
     \centering
     \includegraphics[width=0.9\linewidth]{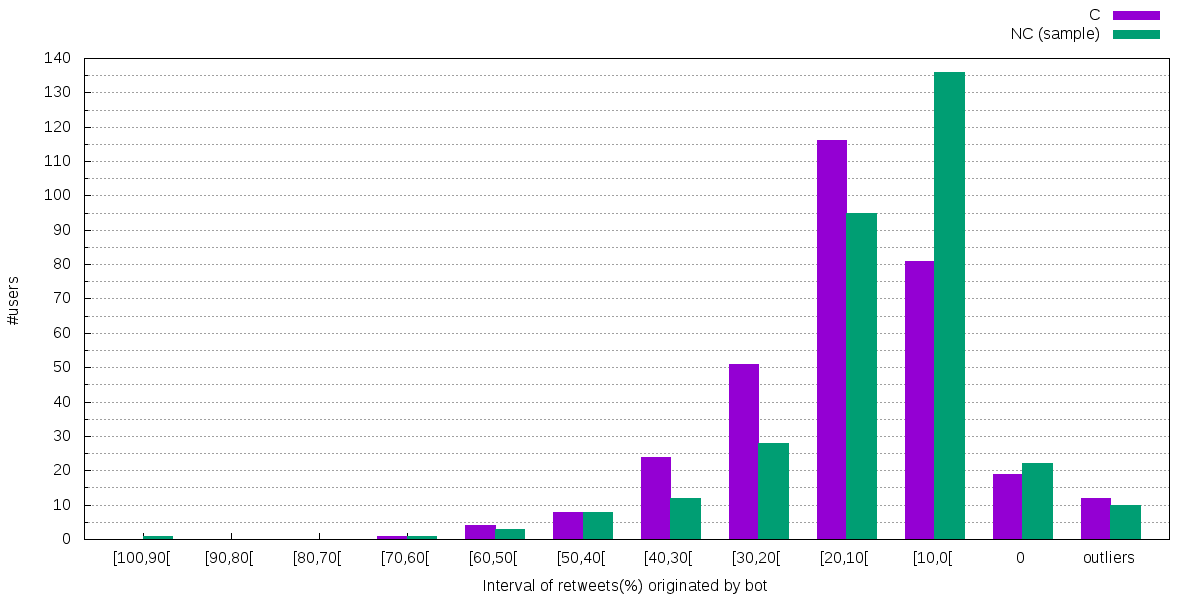} 
     \caption{Deciles of Figure~\ref{fig:rwperc}} 
     \label{fig:rwdec} 
   \end{subfigure}
   
   \begin{subfigure}[b]{1\linewidth}
     \centering
     \includegraphics[width=0.9\linewidth]{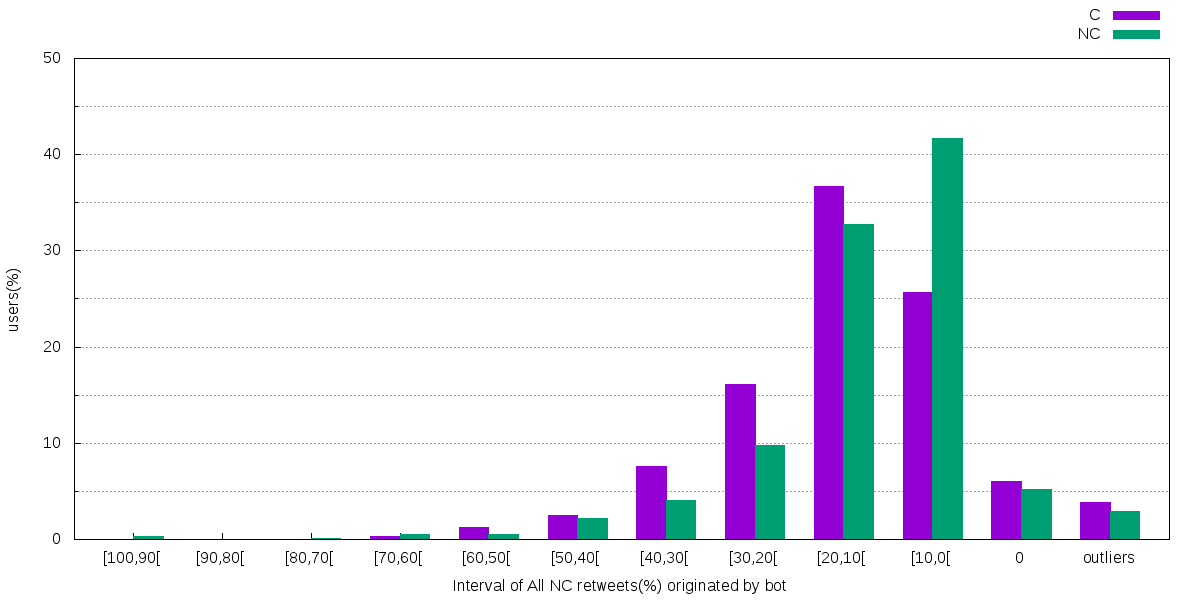} 
     \caption{Deciles of C and all NC users} 
     \label{fig:rwdecAllNC} 
   \end{subfigure}
    \caption{Comparative analysis between \textit{credulous} and \textit{not credulous} users with respect to byBot-retweets.}
    \label{fig:retweetDeciles}
\end{figure}

Going further with the analysis, Figure~\ref{fig:retweetDeciles} provides two aggregation perspectives, by grouping the C and NC users according to the number of their byBot-retweets. 

In Figure~\ref{fig:rwdec}, the x-axis reports intervals (deciles) of byBot-retweets and the y-axis reports the number of users falling in each interval. Since the two sets (C and NC) have the same number of users (316), we prefer to report the real number of users, instead of the percentage, which however is still easily calculable.
The sample of NC users is the same used for the results shown in Figure~\ref{fig:rwperc}. 

Figure~\ref{fig:rwdecAllNC} considers all NC users. Since they are 2,522, we report the percentage (y-axis). 
When considering the whole population of NC users, we can notice 
very similar results, in fact the differences between C and NC users, already observed in Figure~\ref{fig:rwdec}, are here preserved. This can be interpreted as a close relationship between these two sets, i.e., 
the subset of the 316 NC users considered in Figure~\ref{fig:rwdec}, and all NC users.

Finally, in both  subfigures of Figure~\ref{fig:retweetDeciles}, the users in the last group, i.e., the outliers,  do not retweet any tweet; the users in the 0 group are  users retweeting  tweets  originated by human-operated accounts only.


\paragraph{Findings} 
From Figure~\ref{fig:rw}, we can appreciate a difference in users' behaviour between C and NC users. On average, C users feature a higher percentage of retweets whose original tweets have been originated by bots.
The difference between the standard deviation values for the two populations is negligible, indicating a  behavioural similarity between C and NC users (Figure~\ref{fig:rwperc}).


Regarding the analyses shown in Figure~\ref{fig:rwdec}, both the subfigures show a greater presence of  C users  in almost all the deciles; the only relevant difference is for the \textit{[10,0[} group. In this group, NC users are more than C users.  

\subsection{Replies}
\label{subsec:repliesN}

\begin{figure}[ht!] 
\hspace{-4em}
   \begin{subfigure}[b]{0.6\linewidth}
     \centering
     \includegraphics[width=0.95\linewidth]{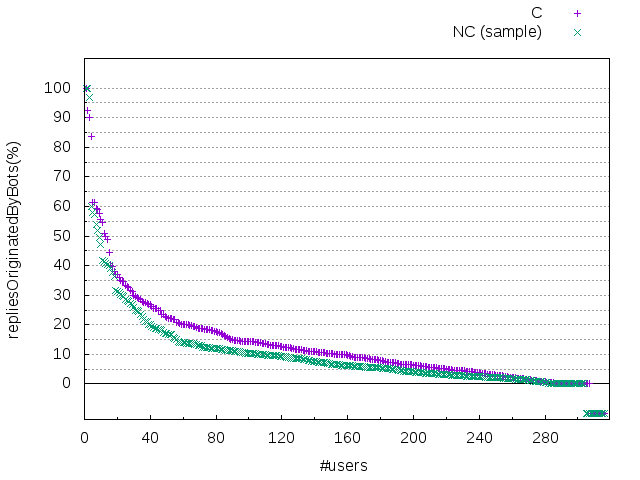} 
     \caption{Percentage of replies to bot's tweets} 
     \label{fig:rperc} 
   \end{subfigure}
   \hspace{-1em}
   \begin{subfigure}[b]{0.6\linewidth}
     \centering
     \includegraphics[width=0.95\linewidth]{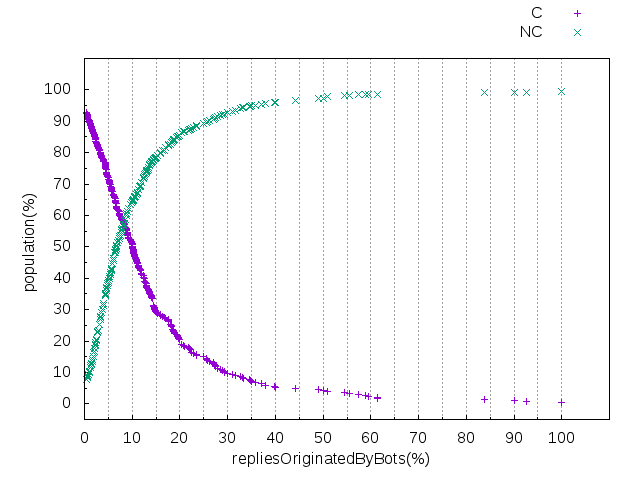} 
     \caption{\% of populations w.r.t. the number of replies to bot's tweets} 
     \label{fig:rcov} 
   \end{subfigure}
   \hfill
    \caption{Comparative analysis between C and NC users with respect to replies to bots' tweets. }
    \label{fig:reply}
\end{figure}

\begin{figure}[ht!] 
\begin{subfigure}[b]{1\linewidth}
     \centering
     \includegraphics[width=0.9\linewidth]{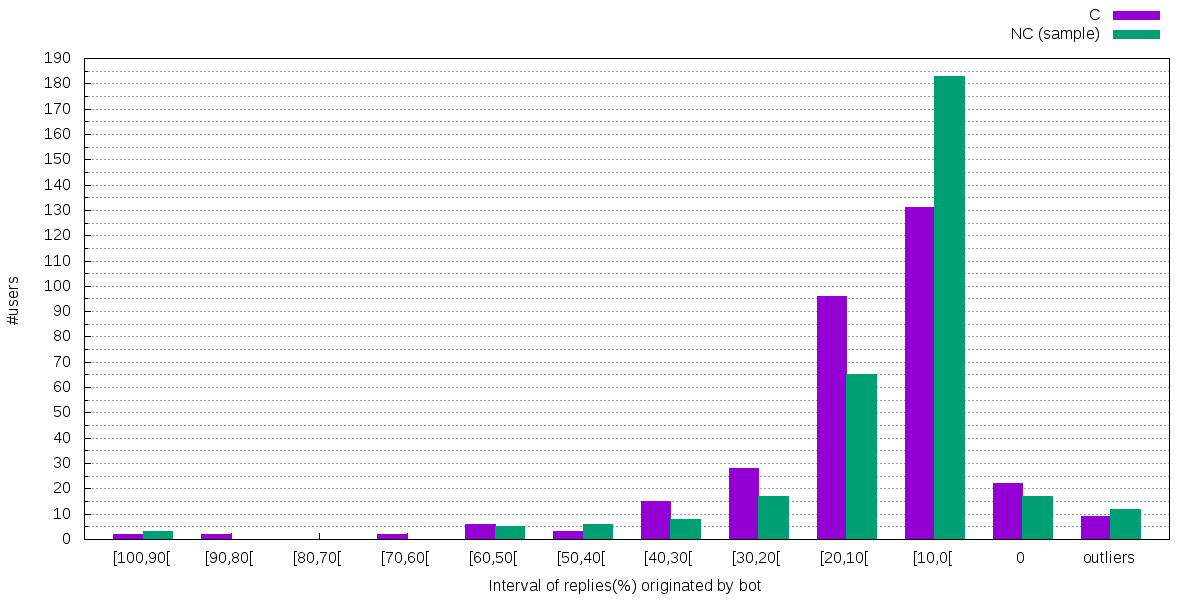} 
     \caption{Deciles of Figure~\ref{fig:rperc}} 
     \label{fig:rdec} 
   \end{subfigure}
   
   \begin{subfigure}[b]{1\linewidth}
     \centering
     \includegraphics[width=0.9\linewidth]{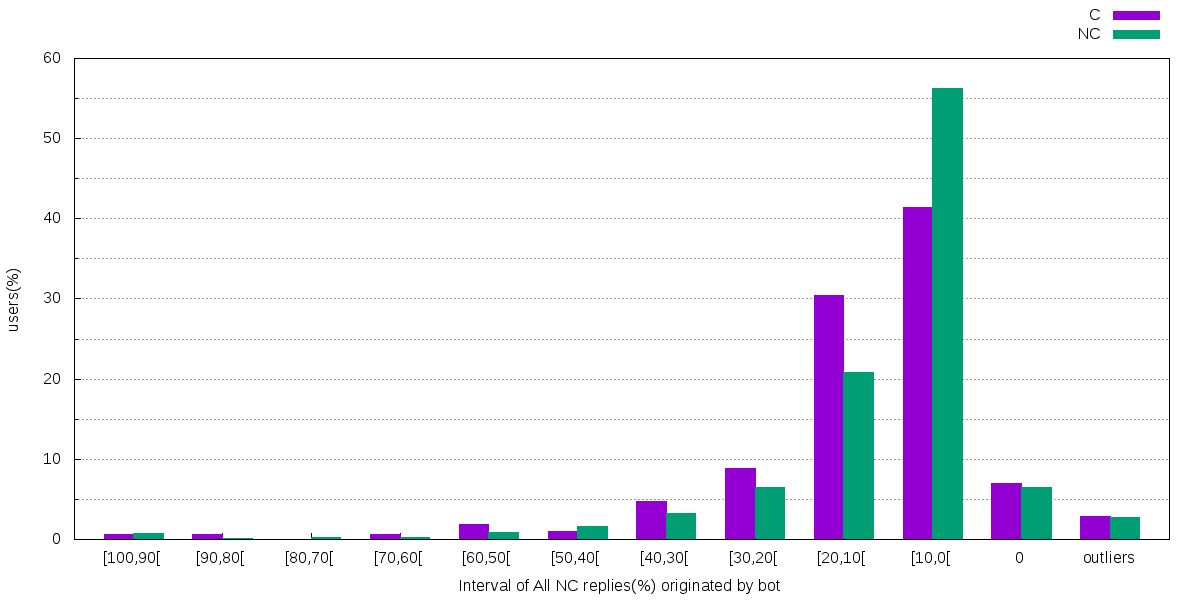} 
     \caption{Deciles of C and all NC users} 
     \label{fig:rdecAllNC} 
   \end{subfigure}
    \caption{Comparative analysis between \textit{credulous} and \textit{not credulous} users with respect to the replies to tweets originated by bots. }
    \label{fig:replyDeciles}
\end{figure}

Figures~\ref{fig:reply} and~\ref{fig:replyDeciles} report the analysis related to the replies.
%
%
Figure~\ref{fig:rperc} shows a quite clear difference between C and (a sample of) NC users. 
C users have an average percentage of replies to bot's tweets equal to 13.77 ($\sigma=15.10$\%), while NC users show a mean's value of 10.81 ( $\sigma=14.03$\%). As for the retweets, the number of outliers is quite low (9 and 12 accounts for C and NC users, resp.).

 Figure~\ref{fig:rcov} shows that the maximum percentage of covered populations is achieved on a replies percentage value equal to 27.96 (x-axis). Specifically, the 11.40\% of C users reply to bot's tweets more than the 91.56\% of NC users. Considering the average percentage value of replies for C users in Figure~\ref{fig:rperc}, 
the populations percentage are 35\% for C users and ~75\% for NC users.

 The behavioral analysis concludes with the bars in Figures~\ref{fig:rdec} and ~\ref{fig:rdecAllNC}.   
The outcomes are very similar to the ones related to the retweets analysis. For most of the groups, there is no a clear distinction between the number in Figure~\ref{fig:rdec} and the percentage in Figure~\ref{fig:rdecAllNC} of  C and NC users. 
This holds at least till the group \textit{[20, 10[}.

\paragraph{Findings} Similarly to what unveiled in the previous subsection, the replies analysis confirms that, on average, C users feature a higher percentage of replies to bots. However, looking  more in detail at the amount of replies (the `group analysis' in Figure~\ref{fig:replyDeciles} , there is no common trend showing, for each group, a majority of C replies w.r.t. NC replies.
To further explore the behavioural difference between C and NC users, we carry out statistical tests, aiming at assessing whether the diversities found up to this point can be considered statistically significant. 

\subsection{Statistical significance of the behavioral differences between C and NC users}
\label{subsec:statsBehavAn}
In Section \ref{subsec:retweetsN} and \ref{subsec:repliesN}, the analyses showed behavioral differences  between C and NC users in terms of retweets and replies. Here, we will try to assess whether these differences can be considered statistically significant. For this purpose, we rely on  hypothesis testing. It is worth noticing that the users involved in the statistical tests, representative of both C and NC users, are those considered in Figure~\ref{fig:rwperc} for retweets and  Figure~\ref{fig:rperc} for replies. 

\begin{table}[htbp]
    \centering
    \begin{tabular}{lcc}
    \hline
& \multicolumn{2}{c}{\makecell{Kolmogorov-Smirnov\\(Test of Normality)}} 
\\\cline{2-3}
& C (Res.)& NC (Res.) \\
\hline
Replies & $\times$ & $\times$ \\
Retweets & $\times$ & $\times$ \\
\hline
\end{tabular} 
\caption{Test of Normality.}
\label{tab:KSTest}
\end{table}

\begin{table}[htbp]
\centering
\begin{tabular}{lccccccc}
\hline
Type of tweets& \multicolumn{3}{c}{\makecell{T-Test\\($\alpha$=0.05)}} & &\multicolumn{3}{c}{\makecell{ANOVA\\($\alpha$=0.05)}}
\\\cline{2-4}\cline{6-8}
& Res. & \textit{t}-value & \textit{p}-value && Res. & \textit{f}-ratio & \textit{p}-value\\
\hline
Replies & $\checkmark$ & 3.001 & 0.001 && $\checkmark$ & 9.04942 & 0.002738 \\
Retweets & $\checkmark$ & 3.190 & 0.001 && $\checkmark$ & 10.17804 & 0.001496\\
\hline
\end{tabular} 
\caption[Parametric Statistical Tests]{Parametric Statistical tests: T-test and one-way ANOVA.}
\label{tab:paramTest}
\end{table}
In Table~\ref{tab:KSTest}, for the two types of post (1st column), we show the results of the Kolmogorov-Smirnov's test~\cite{lilliefors1967kolmogorov} (or \textit{Test of Normality}). This is a non parametric test that, given a certain number of observations (in our case, 
the percentages of retweets and replies of
the two sample populations originated by bots), checks whether they are normally distributed. 
If the test is successful, to determine whether the means of the two sets of data are significantly different we could rely on the outcomes of parametric statistical tests on C and NC users' data, like the T-test~\cite{student1908probable} and the one-way Analysis of Variance~\cite{howell2009statistical} (ANOVA).
%
Unfortunately (see Table~\ref{tab:KSTest}) the two  populations did not pass the test ($\times$ symbol); therefore, the information obtained by the latter mentioned  tests is useless in our situation. For the sake of curiosity and completeness, we considered (see Table~\ref{tab:paramTest}) the outcomes obtained by conducting both the parametric tests on retweets and replies. 
However, we will not consider them further.

\begin{table}[htbp]
    \centering
    \begin{tabular}{lccccccc}
    \hline
Type of tweets& \multicolumn{3}{c}{\makecell{Mann-Whitney\\($\alpha$=0.05)}} && \multicolumn{3}{c}{\makecell{Kruskal–Wallis\\($\alpha$=0.05)}}\\ \cline{2-4}\cline{6-8}
&Res.&\textit{z}-score & \textit{p}-value && Res. & \textit{H}-value & \textit{p}-value\\
\hline
Replies & $\checkmark$ & 3.37056 & 0.00038 && $\checkmark$ & 11.36 & 0.00075\\
Retweets & $\checkmark$ & 3.3 & 0.00048 && $\checkmark$ & 10.89 & 0.00097\\
\hline
\end{tabular} 
\caption{Mann-Whitney and Kruskal-Wallis (non parametric) tests.}
\label{tab:NotParamTest}
\end{table}

We thus rely on non parametric statistical tests. 
In Table~\ref{tab:NotParamTest}, we show the outcomes of two well-known non-parametric tests. The first one corresponds to the non parametric version of the T-test, i.e., test of Mann-Whitney~\cite{mann1947test}. The second one is known as Kruskal–Wallis test~\cite{kruskal1952use}.
For both of them, the test is successfully passed if there is enough grounds to reject the \textit{null} hypothesis. 
Roughly, in both tests, the \textit{null} hypothesis states that ``there is no difference in means'' (of `byBot' content) between the considered populations (in our case C and NC users). 
As we can see in Table~\ref{tab:NotParamTest}, both types of tweets (i.e., replies and retweets) successfully pass the two tests ($\checkmark$ symbol). These results suggest that when replies and retweets are considered, C users interact more with bots than NC users and this behavioural difference is not due to chance.

\medskip

%% file: chapters/5bis-Discussion.tex
\section{Discussion}
\label{sec:discussion}

In Section~\ref{sec:background}, we  presented one of our previous work~\cite{Balestrucci19ideal}, where we:
\begin{enumerate}
\item assessed the capability of a supervised learning-based approach to classify human-operated Twitter users following many bots; \item tested 19 learning algorithms to train a credulous users classifier;
\item evaluated the effectiveness of three sets of features to determine which one obtained  the best classification performances.
\end{enumerate}
Encouraged by promising results (e.g., an accuracy ~93\%) and, therefore, by the ability to automatically identify those users following a large number of bots 
), in this work we extend our studies on C users in an in-depth way.
Specifically, to single out information useful to distinguish C from NC users, we: 
\begin{enumerate}
\item conducted a detailed study on the classification features, by focusing on those used to train our best performing credulous detector (i.e., \textit{ClassA-} features);
\item analysed genuine users' tweeting activities and compare those of credulous with those of not credulous users (a coarse grained analysis not linked to interactions with bots); 
\item conducted a fine grained analysis to check our intuition about the higher engagement of credulous users in spreading content originated by bots.
\end{enumerate}
Regarding  features' analysis, 
we considered three different and well-known feature ranking algorithms and compared them. 
There are small differences in the order in which the features appear in the three rankings. However, 
since the same features appear in the highest positions, we can infer that they are the most effective ones. Some of these high-ranked features are not `Class A' features (i.e., they are not directly accessible from the user profile); indeed, combinations of other features (for example, by division or multiplication) are requested to calculate them. To avoid correlation factors between features, we selected three among the highest-ranked ones that, in our opinion, express the essence of our investigations, namely the number of tweets (a measure of the activity level on the platform), of friends and of followers (a measure of the level of social relationships). For each of these features, we carried out a T-test to assess whether the values associated with C and NC users differ in a statistically significant way. The test succeeded, revealing that these three features unveil a difference between these two populations of human-operated accounts.

Since both C and NC users are human-operated accounts, it is possible that, among the data used to perform meaningfulness tests (on each feature), there may exists some correlations in terms of linear dependency. The statistical test performed on the features (namely F3, F5, F7), although successfully passed, do not take this factor into account. For this reason, we calculated the PCC and found that indeed there is no correlation.
%

 Table~\ref{tab:resumeStats} recaps in a numerical format the statistics of tweeting, retweeting and replying activities of the populations investigated in the previous sections (see Figure~\ref{fig:actions}).

\begin{table}[htbp]
    \centering
    \begin{tabular}{lccccccccc}
         &&  \multicolumn{2}{c}{\textbf{Pure Tweets}}&&  \multicolumn{2}{c}{\textbf{Retweets}}&&  \multicolumn{2}{c}{\textbf{Replies}}\\
         &&$\mu$&$\sigma$&&$\mu$&$\sigma$&&$\mu$&$\sigma$\\
         \hline
         Credulous (C)      && 0.56 & 0.26  &&  0.29 & 0.24 && 0.14 & 0.12\\
         Not Credulous (NC) && 0.46 & 0.25  &&  0.32 & 0.26 && 0.19 & 0.16\\
         NC (sample)        && 0.31 & 0.25  &&  0.31 & 0.25 && 0.18 & 0.16\\
         \hline
    \end{tabular}
    \caption{Tweeting activity (stats) of credulous \textit{vs} not credulous users.
    }
    \label{tab:resumeStats}
\end{table}

On average, C users tweet  more than NC users; nevertheless, their average retweeting and replying activities are lower than those of NCs. At a first sight, credulous users seem more talkative in terms of self-produced content, 
whereas the scenario seems the opposite for retweets and replies. 
Paying more attention,  differences in retweets and replies are not so marked and we can indeed notice similar behaviours of C and NC users. 
This `behavioural equivalence' is exploited in a second and fine-grained behavioural  analysis (Sections~\ref{subsec:retweetsN} and~\ref{subsec:repliesN}). 
Indeed, since the coarse-grained analysis does evidence significant differences between C and NC users, we assume similar behaviour in terms of high level activities (i.e., replies and retweets). The fine-grained analysis enables us to assess the difference in terms of replies to bots, and retweets by bots.

This additional analysis has been conducted both on retweets and replies and has revealed the tendency of C users to bounce more content originated by bots, with respect to NC users. 
To ensure that this behavioural variation does not happen not by chance, we perform further non-parametric statistical tests (hypothesis tests) which confirm the statistical significance of the different attitudes featured by the two categories of users.
We argue that these results provide an initial, but relevant, evidence of the actual  involvement of specific categories of human-operated accounts in supporting, albeit unknowingly, potentially malicious activities.

%% file: chapters/5-RW-shortened.tex
\section{Related work}
\label{sec:RW}


In the following we consider both works that have taken into account the topic of gullibility and approaches that, more in general, consider the behavior of users on social media. We restrict the related work to those papers we consider more relevant relatively to our approach. Thus, this review is not intended to be exhaustive.

For the sake of schematization, Table 
\ref{tb:summary-rw-susceptibility} reports a brief recap 
of the selected papers that are discussed hereafter. 
Our interest is focused on studying users' behavioral patterns, aiming to derive the main characteristics of specific humans, i.e., those more exposed to malicious automated accounts' activities on Twitter, and thus to a higher risk of being influenced by them. 

In a recent study about detection of fake news and mitigation of their harmful effect, Shu and others in~\cite{Shu2019} give a clear definition of different kinds of social media users:  1) the `persuaders', which spread false narratives with supporting opinions to influence others to believe it; 2) the `clarifiers', which propose opposite views to debunk such narratives, and 3) the `gullible users', those open to believe them.
We have investigated the possibility that gullible users are characterized by more interactions with entities such as automated accounts, when compared to non-gullible users. The measure that defines  gullibility of a user is the amount of automated accounts that the user has among her/his friends. 

Individual behaviour in relation to actions and  thinking by other people has been studied in the social sciences for many years. The studies have led to the definition of characteristic aptitudes of the individual, such as the confirmation bias~\cite{confbias98}, i.e., the tendency `to trust information that confirms personal preexisting beliefs or hypotheses',
the cognitive closure~\cite{kruglanski1996}, i.e., the need of obtaining `certainty in an uncertain world', and the selective exposure~\cite{FREEDMAN196557}, i.e., the preference for `information that confirms preexisting attitudes', to mention a few of them.
%
With the advent of internet and social media, the analysis of individual behaviour w.r.t. to communities and their beliefs has been projected by data scientists onto the virtual behaviour of users on the net.  
In order to understand who and what influences users the most, and to what extent they can be influenced, in the recent survey  Zhou and Zafarani~\cite{zafarani2020} devote a small section to what they call `vulnerable normal users'. This designation identifies `users that are susceptible and vulnerable to fake news'. Social theories attest that a reason why a user engages in spreading fake news in a naive way (i.e., without any malicious intent) is that spreading bears a greater  social influence~\cite{Ashforth1989}.


\input{tables/RW-shortened}

Probably, the work most similar to ours is the one by Wagner et al.~\cite{wagner2012social}, dated 2012. In that work, the accounts that here we call credulous are referred to as susceptible. Even in that work susceptible users are successfully recognized by a classifier, but the premises, and the aims of~\cite{wagner2012social} are profoundly different from ours.  
The two works have in common that they do not focus on detecting social bots but on detecting users who are susceptible to their attacks. 
However, there is a substantial difference in the definition of our credulous users and the susceptible users of~\cite{wagner2012social}. A susceptible user is a human that has been `infected' by a bot, i.e., has interacted at least once with a bot, either by mentioning it, retweeting it, or replying to it. 
For us, the credulous user is a user with a large number of bots among her friends. 
The construction of the datasets is also very different. In fact, \cite{wagner2012social} inherits accounts and their interactions from the Social Bot Challenge 2011 - a competition organised by the WebEcologyProject. Thus, Wagner et al. started with a ground truth of genuine bots and accounts, plus a list of their interactions. We also started with datasets of accounts a priori known as genuine ones but then ran a bot detector on their friends to see how many bots they had as  friends~\cite{Balestrucci19ideal}. Here we study whether C users interact with bots differently than NC ones. Finally, Wagner at el. had the goal of discriminating the susceptibility level of the susceptible accounts, a goal that is out of scope here. Moreover, the results of the analysis of the susceptibility level were somehow inconclusive, in the sense that the granularity with which the level of susceptibility was discriminated was very coarse. In light of this, it would be very interesting to understand to what extent it is possible to understand the level of credulity of our credulous users.

A concrete example of the greater exposure of gullible users to deceptive news is given in the recent work by Florendo et al.~\cite{Florendo2019}, which highlights how gullibility is, along with demographic factors, one of the features that have led social media users to believe false news about financial markets. Thus, we think that automatically recognizing gullible users and understanding their intrinsic characteristics is one of the cornerstones to build defences to the spread of false news.


Human reactions are obviously multiple: we do not know `a priori' if  C users approve or not the content they consume and possibly retweet. For instance, Lin et al. in~\cite{LIN2019} tested the perceived trust of a set of users towards one fictitious organization that varied the number of retweets concerning an invented story about contaminated food in grocery stores. 
In this study, a `snob effect' was demonstrated, that is, the more the story was retweeted, the more people tended not to trust the truth of the tweets. 
Other studies  show different reactions. For example, Zubiaga et al. found that online users are more active in sharing unverified rumors than they are in later sharing that these rumors were either debunked or verified~\cite{zubiaga16}. Furthermore, even a bit in disagreement with the previous result, Owen has shown that even after knowing that a story is false, a third of the users continue to spread it anyway~\cite{owen2018}. Overall, it seems that `the veracity of information therefore appears to matter little', as observed by Nemr and Gangware in their report on Weapons of Mass `Distraction'~\cite{gangware2019weapons}. 
Nevertheless, even for the scrupulous reader, it would be very difficult to find out the level of truthfulness of a news, just by using the critical sense. The literature has made progress with the use of automatic tools that exploit the automatic processing of natural language, as demonstrated - for example - in a recent work by Barr{\'{o}}n{-}Cede{\~{n}}o et al. on the detection of propaganda articles~\cite{BARRONCEDENO2019}.

To understand users' behavior on social networks, some crucial points have been identified by Jin et al. in~\cite{jin2013understanding}. Among others,
%
a key aspect is represented by connection and interaction, i.e., the representation of the relationships among users through different types of social graphs, e.g., friendship, following, etc. 
Inspired by this point, our work aims to investigate the behaviour of users related by the Twitter \textit{followees} relationship, since there might be users that are more exposed to malicious activities.

A framework for the detection, estimation, and characterisation of Twitter accounts is presented by Varol et al. in~\cite{varol2017online}, where more than a thousand features are used to discern humans from social bots. When characterising friendship ties and information flow between users, two main findings hold on average, i.e., (i) reciprocity of friendship ties is higher for humans, and (ii) humans resulted to interact more with human-like accounts than bots. As opposite, in this paper we are interested to spot those humans that, maybe unknowingly, diffuse content generated by bots.

The central role of bot accounts in contributing to retweet news 
and amplifying the hubs' messages
has been recently observed in 
Caldarelli et al.~\cite{caldarelli2020squads}. 
%
Given the prominent role, testified by a decade long literature, on the harms that social bots may cause, it becomes of uttermost importance to find out automatic methods to unveil who listens to them, and to what extent. 
Hence, we firmly believe that new approaches should be explored to automatically detect those who heavily interact with the bots.

To the best of our knowledge, most of the literature on social network analysis deals with
detecting bots or assessing the impact of their malicious activities.
%
%
The role of humans, instead, has received less attention, especially when studying misinformation diffusion. Only few attempts have been made to identify those social media human users that are susceptible to disinformation attacks by social bots. 
%
%
Users that are most vulnerable to social bots 
were considered 
in~\cite{wald2013predicting}, where Wald et al. conducted some experiments to derive the characteristics of users replying
to bots or following them. 
From their experiments emerged that the Klout score\footnote{Klout is a private company collecting information on users acting in different social media (Facebook, Twitter, G+, LinkedIn), to determine their overall social influence.}, the number of friends and followers are the best indicators (among a set of 13 features) to predict whether a human will interact with a bot.
Our work can be considered as complementary to~\cite{wald2013predicting}, in fact we also consider the total number of bots' followees for spotting credulous users. 

Users' retweeting is investigated by Lim and Hoang in~\cite{hoang2013retweeting}, and it is associated to three behavioral factors: (i) topic virality, i.e., the ability of a topic to attract retweets, (ii) user virality, i.e., the ability of a user to get retweeted for a specific topic, and (iii) user susceptibility, i.e., the ability of a user to retweet for a specific topic. In this paper we are mainly interested to retweets induced by user susceptibility, and from~\cite{hoang2013retweeting} we learnt that a small group of users is extremely susceptible to election-related influences. 


Virality and susceptibility in social media is tackled by Hoang and Lim in~\cite{hoang2016tracking}, the focus being on the temporal dynamics of the behavioral factors that were neglected by the same authors in~\cite{hoang2013retweeting}. Time models are proposed to assign higher/lower susceptibility score to users on the basis of 
retweeting activities during specific time steps. 
Our work also 
does not consider the temporal aspect to lighten the computational cost. However, as future work we plan to study how the behavior of credulous users change over time. 


More recently, there has been some research effort devoted to detecting users susceptible to fake news. In~\cite{DBLP:conf/websci/ShenCGLYL19}, Shen et al. start from a dataset of fake news, and all the Twitter users replying to such news are labelled as vulnerable to disinformation. A supervised classification is later adopted to train a model that classifies gullible users, according to content-, user-, and network-based features. Results show the capability to differentiate users with different susceptibility levels, achieving 0.82 in AUC-ROC as best performance value. 
Also in this paper we analyse the content originated by bots and disseminated by human users. 
In particular, we study how potentially fake content (because originated by bots) are disseminated by credulous users who, although unknowingly, can actively contribute to the dissemination of fake news.

A framework to identify polarised content on social media and to predict future fake news topics is proposed Del Vicario et al.~\cite{vicario2019polarization} which  use a number of characteristics related to users behavior (e.g., number of likes, comments, and shares) for the classification task. 
It would be interesting to design ad-hoc experiments to exploit these characteristics by leveraging those values that are associated to potential targets for hoaxes and fake news. This way, we can detect users that are susceptible to and potential contributors of misinformation spreading.

The influence of fake news in Twitter has been examined in~\cite{bovet2019influence} where Bovet and Makse analyze the information related to the 2016 US presidential election. Results of this study demonstrate that Clinton supporters were largely influenced by the spreading of center and left leaning news,
whereas Trump supporters were heavily influenced by the dynamics of the top fake news spreaders. 
Similarly to approaches on fake news~\cite{DBLP:conf/websci/ShenCGLYL19,vicario2019polarization,bovet2019influence}, our interest is on verifying if 
users contributing to spreading of fake content are among our credulous users.

%% file: tables/RW-shortened.tex
\begin{table*}[t]
\centering
\caption{Summary of the most relevant related work.}
\scriptsize
\label{tb:summary-rw-susceptibility}
\begin{tabular}{ c | p{14.5cm} 
}
\hline
\multirow{2}{*}{\textbf{\emph{Ref.}}} & \multirow{2}{*}{\textbf{\emph{Brief summary}}} 
\\
& \\ \hline 
\multirow{2}{*}{\cite{Shu2019} } & \multirow{2}{*}{taxonomy of social users according to susceptibility, persuasion, and aptitude to clarification levels} \\
  &  
\\ \hline 
\multirow{2}{*}{\cite{Florendo2019}} & \multirow{2}{*}{inclination of susceptible users to listen to fake news regarding financial markets} \\
  &  
\\ \hline
\multirow{2}{*}{\cite{LIN2019}} & \multirow{2}{*}{study on the perceived trust of social users towards massive retweet campaigns}\\
  & 
\\ \hline
\multirow{2}{*}{\cite{zubiaga16}} & \multirow{2}{*}{social users' aptitude to share unverified rumours}\\
  & 
\\ \hline
\multirow{2}{*}{\cite{owen2018}} & \multirow{2}{*}{persistence of social users to share rumours even if debunked or verified}\\
  & 
\\ \hline
\multirow{2}{*}{\cite{gangware2019weapons}} & \multirow{2}{*}{reasoned motivations that lead social users to believe and spread unverified news}\\
  & 
\\ \hline
\multirow{2}{*}{\cite{BARRONCEDENO2019}} & \multirow{2}{*}{propaganda and how to spot it in online news} \\
  & 
\\ \hline
\multirow{2}{*}{~\cite{jin2013understanding}} & \multirow{2}{*}{users' behaviour on social networks is influenced by connections and interactions} \\
  & 
  \\ \hline
\multirow{2}{*}{~\cite{varol2017online}} & \multirow{2}{*}{characterisation of Twitter accounts with features discerning humans from social bots} \\
  & 
\\ \hline
\multirow{2}{*}{\cite{caldarelli2020squads}} & \multirow{2}{*}{strategic formation of bots squads to amplify political messages on Twitter}\\
  & 
\\ \hline
\multirow{2}{*}{\cite{wagner2012social}}  & \multirow{2}{*}{susceptibility of human users quantified in terms of interactions, i.e., mentions, replies, retweets, etc.}
 \\
   & 
\\ \hline
\multirow{2}{*}{\cite{wald2013predicting}}  & \multirow{2}{*}{studying the characteristics of users replying to or following bots} 
\\
  & 
\\ \hline
\multirow{2}{*}{\cite{hoang2013retweeting}} & \multirow{2}{*}{investigation of users' retweeting 
to understand the features of susceptible users attracted by election topics} \\
  &  \\ 
\hline
\multirow{2}{*}{\cite{hoang2016tracking}}  & \multirow{2}{*}{tracking susceptibility in social media by considering the temporal dynamics of the behavioural factors} \\
  &  \\ \hline
\multirow{2}{*}{\cite{DBLP:conf/websci/ShenCGLYL19}}  & \multirow{2}{*}{building a dataset of Twitter fake news followers by selecting all the accounts replying to known fake news} \\ 
  & 
\\ \hline
\multirow{2}{*}{\cite{vicario2019polarization}} & \multirow{2}{*}{identifying polarised content on social media (based on users behaviour) and predicting future fake news topics} 
\\
  & 
\\ \hline
 \multirow{2}{*}{\cite{bovet2019influence}} & \multirow{2}{*}{fake news spreaders in Twitter during the US presidential election influenced Trump supporters} \\
  &  \\
\hline
\end{tabular}
\end{table*}

%% file: chapters/6-Concl.tex
\section{Conclusion}
\label{sec:concl}
Disinformation spreading on social media is a worrisome phenomenon to which researchers, platform administrators, and even governments are looking at with concern. The role of bot accounts in this business is unquestionable, but it would not be effective if there was nobody considering them. The work presented in this paper aimed precisely to test the attitude of human-operated accounts towards reacting to the actions of bots. 
To this purpose, 
we have considered Twitter online accounts which have a high number of bots among their friends; we have named them as credulous users. Leveraging a classification work carried out in  our previous work, we have analysed the statistical value of the features considered for such classification phase. Such analysis has enabled us to conclude that some features, such as the number of tweets, of friends and of followers, that can be easily extracted from the account's profile, are statistically relevant to discriminate between Credulous and Non Credulous users. 


Besides, by considering the retweets and the replies of the accounts in our datasets, we have shown, through two statistical tests, that, on average, C users amplify more than NC ones the content posted by bots. Even before conducting further experimental analysis on larger samples of C users, we consider this result very promising. Indeed, it shows that it is possible:
\begin{enumerate}
    \item to automatically identify credulous users accounts by leveraging on discriminating features that are very easy to extract;
    \item to get useful information on possible dissemination of spam content, propaganda, and, in general, of unreliable information, by focusing on the source of the content credulous users bounce.
\end{enumerate} 

\paragraph{Future Work}

Despite these encouraging results, we argue that scholars and platform admins should put  more effort to make users aware of the pitfalls that can be hidden by interacting with accounts, let them be automated or not, whose purposes are not honest.
Hereafter, we propose some possible future investigations:
\begin{itemize}
    \item [-] observe the variations of credulous users' followees and check, by considering an observation time frame, the nature (genuine vs bots) of those who have started to be followed, those who have stopped being followed and those who stay longer on the followees lists.
    This study could help understanding the constancy of a C user in being susceptible to possibly not reputable content.
    \item [-] develop approaches for C users detection also for human-operated accounts with more than 400 followees.
    Investigations in this direction would further contribute to understanding whether the proportion of suspicious users that a C user follows is proportional to the number of followees.
    \item[-] adapt the approach to other social platforms. The concept of C users is strongly dependent on the specific relationships between users on the specific social platform, thus the concept of being interested in published content deserves specific attention.
\end{itemize}

\section*{Acknowledgements}
Partially supported by the European Union's Horizon 2020 programme (grant agreement No. 830892, SPARTA) and by IMT Scuola Alti Studi Lucca: Integrated Activity Project TOFFEe `TOols for Fighting FakEs'.
It has also benefited from the computing resources (ULITE) provided by the IT division of LNGS in L'Aquila.